\documentclass{sigchi}

\usepackage{balance}       
\usepackage{graphics}      
\usepackage[T1]{fontenc}   
\usepackage{txfonts}
\usepackage{mathptmx}
\usepackage{color}
\usepackage{booktabs}
\usepackage{textcomp}
\usepackage{graphicx}
\usepackage{subcaption}

\usepackage{microtype}        
\usepackage{ccicons}          

\usepackage{todonotes}

\def\plaintitle{A New Terrain in HCI: Emotion Recognition Interface \\ using Biometric Data for an Immersive VR Experience}

\def\plainkeywords{emotion recognition interface; virtual reality; seamless sensing system; EEG; eye-tracking}

\makeatletter
\def\url@leostyle{%
  \@ifundefined{selectfont}{
    \def\UrlFont{\sf}
  }{
    \def\UrlFont{\small\bf\ttfamily}
  }}
\makeatother
\def\pprw{8.5in}
\def\pprh{11in}

\definecolor{linkColor}{RGB}{6,125,233}

\begin{document}

\title{\plaintitle}

\numberofauthors{4}
\author{Jaehyun Nam \\ \mdseries{Looxid Labs} \\\and
Hyesun Chung \\ \mdseries{Seoul National University} \\\and
Young ah Seong \thanks{Contact authour: yabird@gmail.com} \\ \mdseries{The University of Tokyo}  \\\and
Honggu Lee \\ \mdseries{Looxid Labs} \\
}

\maketitle

\begin{abstract}
  Emotion recognition technology is crucial in providing a personalized user experience. It is especially important in virtual reality(VR) to assess the user's emotions to enhance their sense of immersion. We propose an emotion recognition interface that incorporates the user's biometric data with machine learning technology for increasing user engagement in VR. Our key technologies include brainwave sensors and eye-tracking cameras embedded in a VR headset, which seamlessly acquire physiological signals, and secondly, an attractiveness recognition algorithm that uses bio-signals to predict the user's attraction on visual stimuli. We conducted experiments to test the performance of the system, and also interviewed experts and participants to acquire opinions on the system. This study demonstrated the technical feasibility of our system with high accuracy and usability. Interviewees expected that the interface will be actively used in the context of various applications. Our proposed interface could contribute to an immersive VR experience design.
\end{abstract}

\ccsdesc[500]{Human-centered computing~Human computer interaction (HCI)}
\ccsdesc[300]{Human-centered computing~Virtual reality}
\ccsdesc[100]{Human-centered computing~User studies}

\keywords{\plainkeywords}

\printccsdesc

\section{Introduction}
A key component of human-computer interaction (HCI) is the ability of a computer to understand scenes or items to which a person finds appealing. Currently, the most widely used approach that is employed to achieve this is by giving an explicit rating, which is a self-assessment method in which a person directly evaluates a product or a service. Because most people do not explicitly appraise content as they are viewed, recently there have been efforts to develop implicit rating methods to infer persons' assessments~\cite{choi2012hybrid,claypool2001implicit,schein2002methods}. Through advances in big-data and machine learning technologies, users' assessments of attractiveness can be predicted using a variety of data without having to refer to the users. Using log data, users' interests are obtained about the main information that they access, and their personal preferences are analyzed by tracking major activity areas using the global positioning system (GPS) feature of their smartphones~\cite{liu2010personalized,park2007location,ghose2011empirical}. The analyzed information enables the interaction to recommend products or services in which they would be more interested.

Various emotion recognition technologies have been developed to understand human emotion with less conscious processing. Facial recognition analyzes emotions using human facial expressions, and voice recognition analyzes them using various voice features such as intonation, volume, and speed~\cite{yu2015image,happy2015automatic,zhang2016facial,mollahosseini2016going,wang2015speech,anagnostopoulos2015features}. Text mining is another emotion recognition technology that utilizes the words or context used in writing~\cite{yadollahi2017current}. Because data such as facial expressions, voice and text are types of information that can be collected naturally during interactions between humans and computers, not only is the data collection process simple, but the use of analytical information is also valuable. However, these data still comprise filtered information according to the situation or environment rather than raw expression of human emotion~\cite{smith2015neural}.

In recent times, an emotion-recognition method that utilizes bio-sensing data has attracted newfound attention. Bio-signals show excellent data consistency for emotion analysis because they represent unfiltered and immediate responses to human emotions. The emotional status can be analyzed by directly monitoring their brain activity using functional magnetic resonance imaging (fMRI) or electroencephalogram (EEG), as well as biometric data including galvanic skin response (GSR), photo-plethysmography (PPG) and electromyography (EMG)~\cite{vecchiato2010changes,yadava2017analysis,paulus2003ventromedial,bar2007visual}. It is important that the biometric data be acquired naturally in the real world for the technology to be widely accessible.

Head-mounted displays (HMDs) are an interesting digital platform for the virtual reality (VR) environment. HMDs are optimal digital platforms for easily and naturally collecting biometric data because they have direct contact with individuals' faces. In particular, the forehead area on which the head-foam cushion makes physical contact is in front of the area of the brain responsible for human cognition and emotion, namely the prefrontal cortex, so it is possible to obtain information on changing brain activity according to different emotions~\cite{cardinal2002emotion,etkin2011emotional}. Moreover, it is more important to understand a user's emotions in a VR environment than in other digital platforms. In order to make the VR more real, HCI that is based on emotional status rather than conventional I/O interaction is necessary~\cite{schuemie2001research}.

In this paper, we propose a new emotion recognition interface that predicts users' evaluations of attractiveness using biometric data acquired from a self-developed VR sensing system. We focus on attraction or preference only in particular, as a stepping stone to see if inferring other human emotions from biometric data can be possible. We designed and built a VR accessory with EEG sensors and eye-tracking cameras embedded into a VR headset. The proposed system seamlessly measure users' brain activity and eye movement in a VR environment. Then, we developed a machine learning model that analyzes attractiveness using biometric data, and we conducted experiments to verify the feasibility of the system. Finally, we considered the prospects, advantages and limitations of how our proposed system will contribute to the development of HCI in the future by performing interviews with participants and experts in various fields including VR, AI, and brain engineering. We considered:

A new emotion recognition interface in VR that allows a seamless measurement and prediction of the user's attraction state by the use of unobtrusive EEG sensors and eye-tracking technology.

A guideline for how the interface contribute to enchiching user experience in VR applications.

\section{RELATED WORK}
\subsection{Emotion detection using biometric data in HCI}
With the development of wearable EEG headsets, studies on the brain-computer interface (BCI) have begun to vary outside the laboratory environment~\cite{liu2010real,atkinson2016improving}. In particular, advances in EEG wearability have significantly assisted HCI researchers. In Nick et al~\cite{merrill2018scanning}'s study, developers in Silicon Valley proposed that the continued evolution of BCI technology would enable interactions that involve the reading users' minds to be fully accepted and utilized by users. However, EEG headsets have fewer sensors than conventional EEGs, so the amount of data is smaller, and data quality from the headset is poor owing to the use of dry electrodes. Numerous studies have attempted to achieve higher accuracy in this environment, but there are clear limitations with respect to the system configuration.

There are various research directions that can remedy the limitations of the BCI system. The first method is a multi-modal system that uses a variety of biological signals simultaneously to increase the amount of information. If bio-signals that have already been proven to be correlated with emotion, such as the heart rate or GSR, are used in the study together with, they can compensate for the lack of information in the BCI system~\cite{vecchiato2010changes,soleymani2012multimodal}. In a multi-modal system, it is important to increase the analysis accuracy by controlling each signal from the different modality at different sampling rates. Other studies are attempting to improve accuracy using machine learning techniques. Various methods such as texts, voice, and facial expressions have been proposed as the most prominent machine learning techniques in emotion recognition technology~\cite{cowie2001emotion, trigeorgis2016adieu, mistry2017micro, li2017targeting}. There are increasing number of studies that apply machine learning techniques to biometric data. However, there is no set of classification for raw bio-signals the way facial expressions do for their corresponding emotions. Based on these technical considerations, we designed our attractiveness prediction algorithm in accordance.

\subsection{Importance of emotion detection in VR}
In the new digital platform of VR, it is important for users to feel that they are actually present in the environment. Sense of presence is greatly influenced by how naturally a user interacts with the character or environment in the content. In particular, because a VR user exists as a participant in the three-dimensional (3D) world and not as a third-person observer with a limited perspective like in other digital platforms, the interaction of users and content in VR is much more important than elsewhere.

In order to design how interactions between users and content should occur, we need to understand interactions with each other in the real world. According to Albert's study~\cite{mehrabian2009silent}, the three most important elements of message delivery in communication are suggested to be word, tone, and body language, which account for 7\%, 38\% and 55\% of messages, respectively. (the 7\%-38\%-55\% rule) Words with explicit meaning have only a 7\% significance, while interactions with implicit language have an impact of 93\%. That is, communicating via emotional interaction rather than through explicit I/O systems is a much more important component of immersive interaction. This principle should also be applied to the HCI method in VR environments.

Early studies on VR interaction have focused on increasing the sense of immersion by providing users with sensory feedback such as visual and haptic feedback, or by developing an interaction method based on physical activity such as motion capture and treadmills~\cite{lecuyer2010using}. Recently, researchers have recognized the importance of emotion-based interaction, and several attempts are being made to achieve this. In addition to utilizing the mental activity of users through BCI technology~\cite{abdessalem2017real,guo2013new}, they are also attempting to form an emotional connection between the user and characters by way of content-based techniques such as creating eye-contact and animal sniffing them. VR content companies proposed that while a physical I/O system is also important for immersion, the emotional connection with the characters is when users feel the sense of presence. However, because the content cannot take into account the user's emotions continuously, there is an increased need for a more technical approach. Keeping this in mind, we would like to examine in this study aims to examine how our systems can contribute to HCI in the VR field.

\section{EMOTION RECOGNITION INTERFACE}
\subsection{System overview}
Our proposed system as shown in Figure~\ref{fig1}, which consists of a VR headset with embedded EEG sensors and eye-tracking cameras, as well as time-synchronizing data acquisition system, was designed to achieve two main goals. First, we tried to develop a framework to seamlessly and naturally measure the user's physiological signals when it looks at various contents in a virtual environment. Second, the system aims to easily find and analyze EEG- or eye movement-related data of the exact moment in which researchers are interested using time synchronization~\cite{eidson2002ieee} between different bio-signals and events.

In general, EEG devices and most of the other prevalent biometric equipment are too bulky and cumbersome to wear. Therefore, the use of those devices provides users with unpleasant experience with respect to wearing it and following the directions in an experimental setting. To address the usability issue, we embedded a light-weight, dry-electrode EEG sensor in the HTC Vive headset, especially where the headset and the user's forehead touch each other. In addition, we installed cameras in the lower part of the lens to measure the user's pupil dilation and eye movement. With this transformed hardware interface, the users' brainwave and eye movement data can be easily acquired while they ordinarily wear HMDs. 

\subsubsection{Sensors}
\begin{figure}[!t]
\includegraphics[width=0.45\textwidth]{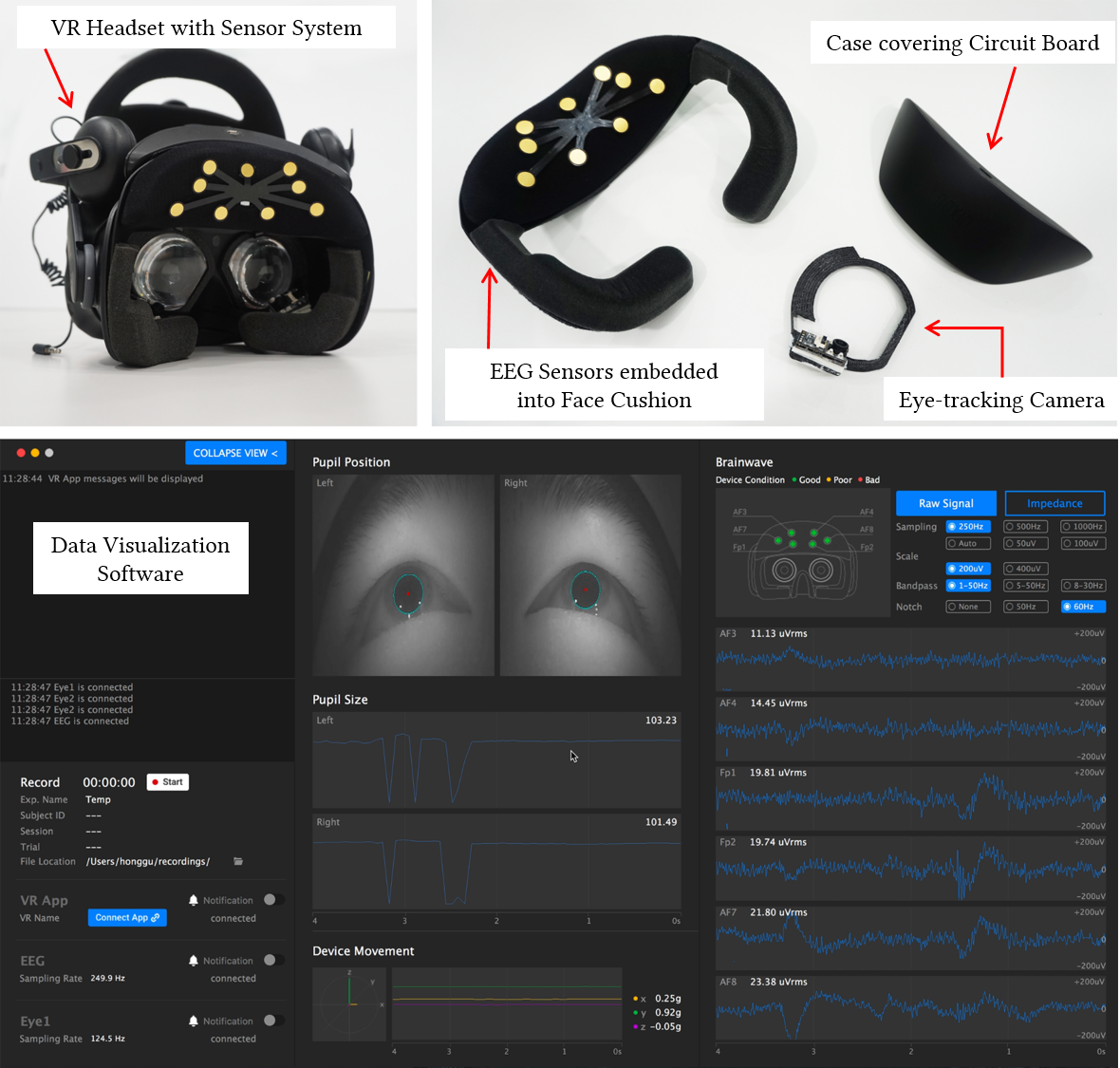}
\caption{HTC Vive headset equipped with a seamless sensing system.}
\label{fig1}
\end{figure}

Based on the neuroscience research that the brain's prefrontal region contains information in that deals with emotion~\cite{badre2004selection, wallis2003neuronal, tsujimoto2006direct, ariely2010neuromarketing, boksem2015brain}, we have located EEG sensors on the user's forehead corresponding to this area. The specific points (Fp1, Fp2, AF3, AF4, AF7, and AF8) were selected following the International 10-20 system~\cite{homan1987cerebral}. The 1mm-thick EEG sensor is tin plated with gold, and is printed on a flexible printed circuit board (FPCB)(Figure~\ref{fig1}). The greatest advantage of this approach is the high flexibility that allows a comfortable fit regardless of different forehead shapes. The raw EEG signals are recorded at a sampling rate of 250 or 500 Hz.

Two eye-tracking cameras are located at 45 degree below the horizontal line of the users' eyes when looking forward(Figure~\ref{fig1}). The angle was calculated in order to obtain a clear recording of each eye's contour, and to achieve a suitable pupil detection function. The eye movement is stored both in a video format and as time-variant feature data, at a sampling rate of 60 Hz. The acquired gaze information is mainly used to track what the user exactly stares at inside a virtual environment. Other important data include the dilation and moving speed of the pupil which can be used to infer several emotions such as stress or fear~\cite{van2015prolonged}. The specifications of each sensor are described in Table~\ref{lxvr_spec}.

\subsubsection{Time Synchronization}
When analyzing biometric data, it is very important to know the exact times at which the stimuli occur, and to match them to the corresponding bio-signals~\cite{busch2009phase}. Each sensor that is included in our system has a different sampling rate, and events that represent the change of contents or the user's expected reaction always happen arbitrarily. Therefore, in this multi-modal sensing environment, it is critical to synchronize the acquired data according to the events. Our system has realized it well such that the output brainwave and eye movement data are arranged according to the time-stamp of any single event.

\begin{table*}
\centering
  \caption{Specifications of EEG sensor and eye-tracking camera in the system.}
  \label{lxvr_spec}
  \begin{tabular}{cccc}
    \toprule
    \multicolumn{2}{c}{EEG Sensor}&\multicolumn{2}{c}{Eye-tracking Camera}\\
    \midrule
    Number of channels &	6	&	Number of channels	&	2\\
    Location of channels &	Fp1, Fp2, AF3, AF4, AF7, AF8 & Location of channels	& Left and right eye \\
    Sampling rate & 250, 500 Hz& Sampling rate& 60 Hz\\
    Resolution & 24bit & Resolution & 640*400 \\
    CMRR & > 100dB & Gaze coordination accuracy & 0.3 degree \\
    Input-referred noise & < 1.5 uVpp (EXT) & Data output & Relative pupil position,\\
    & & &  size and gaze coordination \\
  \bottomrule
\end{tabular}
\end{table*}

\subsection{Attractiveness recognition model}
\begin{figure}[!t]
\includegraphics[width=0.45\textwidth, keepaspectratio]{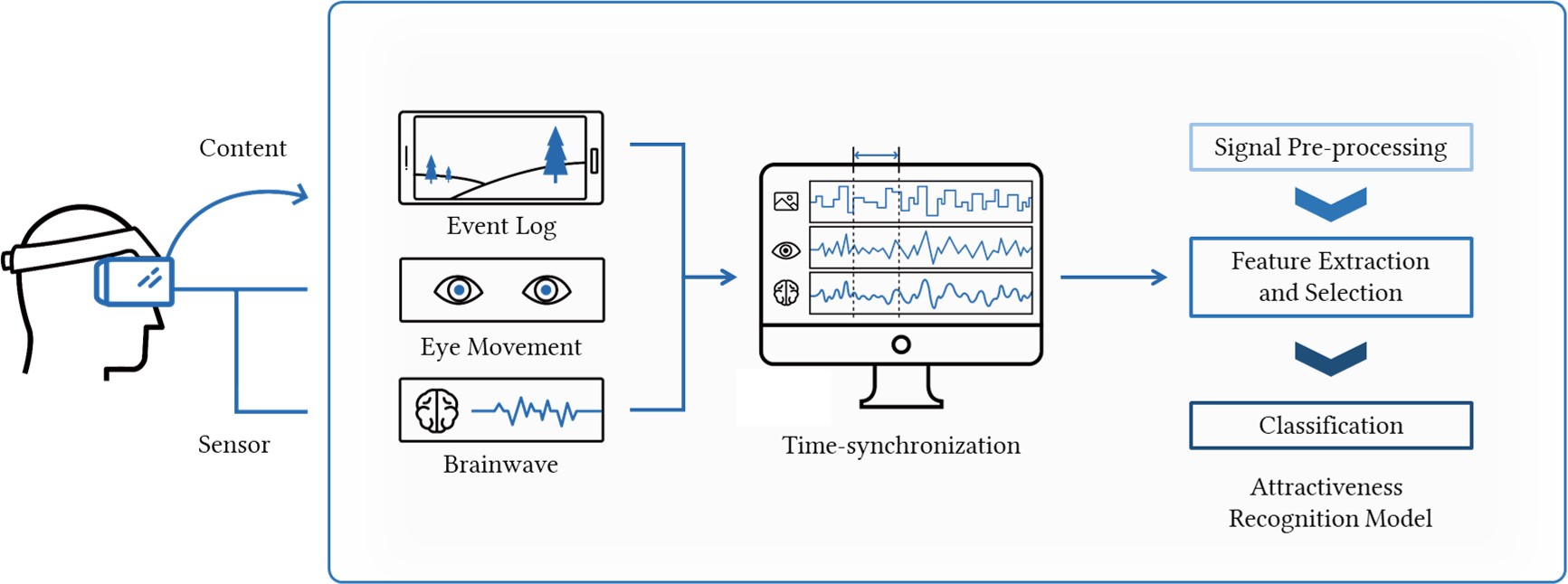}
\centering
\caption{Emotion recognition interface framework.}
\label{fig_system}
\end{figure}

In this work, we designed a multi-modal attractiveness recognition framework that enables the recognition of the user's attraction using brainwave and eye movement data that are acquired from the hardware described above. As attraction is a highly complicated and personal emotion, a single feature from a single modality is insufficient to correctly evaluate it. Besides, the influence of the features varies depending on the user. Therefore, we have developed technology that improves the accuracy of attractiveness recognition by processing numerous features from the multi-modal system and designing the machine learning algorithm (preprocessing - feature extraction and selection - classification) to facilitate an individualized analysis (Figure~\ref{fig_system}).

\subsubsection{Preprocessing}
In general, the electrical signals from brainwaves are sensitive to noise such as eye movement or interference with electrical devices. Therefore, in order to reduce it and improve the quality of the signal, the raw EEG data go through two preprocessing steps. First, EEG data are digitally filtered with a band-pass filter between 0.01 and 120 Hz, and a stop filter at 60 Hz to reduce power line noise. Then, by independent component analysis (ICA) and applying a modified brain wavelet method~\cite{PatelAX2016}, ocular artifacts and headset slip artifacts are removed. In fact, this preprocessing significantly helps to remove EEG noise caused by user movement.

With respect of eye-tracking data, the pupil diameter is not only determined by the user's emotion but also by luminance. Therefore, preprocessing is also required to properly extract emotion-related features. We applied the pupillary light reflex removal method proposed in~\cite{soleymani2012multimodal}.

\subsubsection{Feature extraction and selection}
After preprocessing, the system extracts numerous features that are known to be relevant to brainwave analysis. In the case of EEG, five primary features are extracted, and there are about 300 feature dimensions. The features consist of time-domain features (non-stationary index (NSI), fractal dimension (FD), higher-order crossings (HOCs)), frequency-domain features (five frequency bands: delta (1-4 Hz), theta (4-8 Hz), alpha (8-14 Hz), beta (14-31 Hz), and gamma (31-50 Hz)), and time-frequency-domain features (discrete wavelet transform (DWT)). In the case of the eye data, mainly five features are used, and there are totally 12 feature dimensions. This includes the mean, standard deviation, power spectral density (PSD) of four frequency bands (0-0.2 Hz, 0.2-0.4 Hz, 0.4-0.6 Hz, and 0.6-1 Hz), fixation duration, and frequency calculated according to the pupil dilation.

However, even though each feature described above contains important information, it is inefficient to apply all of them to the machine learning algorithm because of the computational workload, and the somewhat distract  learning owing to high redundancy. Therefore, only the most relevant features are required to properly measure the user's attraction level. Our system applies the infinite latent feature selection (ILFS)~\cite{roffo2017infinite} for feature selection.

\subsubsection{Classification} 
In order to perform a attractiveness classification, the polynomial kernel function is used in standard support vector machine(SVM), which is optimal when classifying data with a high dimensional feature space. To enhance the reliability of learning and avoid over-fitting, we applied 10-fold cross validation method. That is, dividing the entire feature dataset into ten subsets, where an SVM classifier learns nine subsets and uses the remaining subset as a test set. We aim to test the whole dataset, and thus the process will be iterated for ten times. Finally, the accuracy of the attractiveness recognition algorithm is calculated as the average of each test set's accuracy.

\section{EXPERIMENT}
We conducted two kinds of experiments to demonstrate the technical feasibility of our proposed attractiveness recognition system.  First, to demonstrate the performance of the systems, we performed various analyses using biometric data obtained when participants presented different types of images. Second, we constructed an application that could show participants the results of the attractiveness recognition in real time to receive their feedback on an interface understanding human emotion.

In this study, 13 healthy participants (seven males and six females), who were aged between 22 and 34 years, were recruited to take part in the experiment. They wore HTC Vive equipped with sensors, and viewed various kinds of images. The system automatically stores the users' brainwave and eye movement data. First, the experiments started with brainwave and eye-movement calibration in order to set the baseline for their individual bio-signals. After a calibration session, they completed two different sessions and then filled out surveys on the experiment.

\begin{figure}[!t]
\includegraphics[width=0.45\textwidth, keepaspectratio]{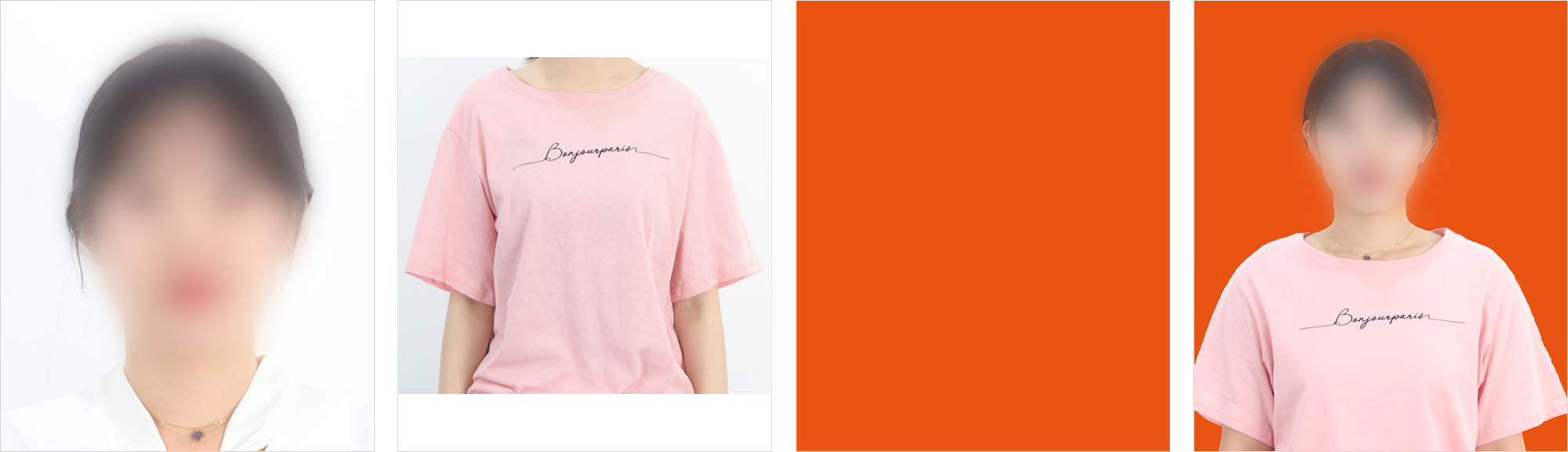}
\centering
\caption{Image of face, clothes, color and a combination of the three.
}
\label{fig_face_color}
\end{figure}

For the first session, we prepared 144 images with four different categories: human face with neutral expression (30), clothes (30), color (30) and a combination of the three (54). There are mainly two reasons for the image-set selection: as most of interactive VR contents include human-related stimuli, we tried to use it to verify the system; we aimed to check whether we can properly observe the degree to which the user appreciates each image containing various elements, and thus prepared the composite image set comprising different combinations of the face, clothes, and color, as shown in Figure~\ref{fig_face_color}. A number of images of faces were first collected by taking upper body photos of 30 models (15 males and 15 females) who we recruited, and who agreed on the use of their photos for the experiment. We created 54 composite images by selecting three images of faces, three images of clothes, and three images of colors for men and women, respectively, from among a total of 90 images of faces, clothes, and colors. Each participant was instructed to stare at the image for 2 s, and had to give a rating on a discrete seven-point scale to indicate the degree of attraction to the image.
\begin{figure}[!t]
\includegraphics[width=0.45\textwidth, keepaspectratio]{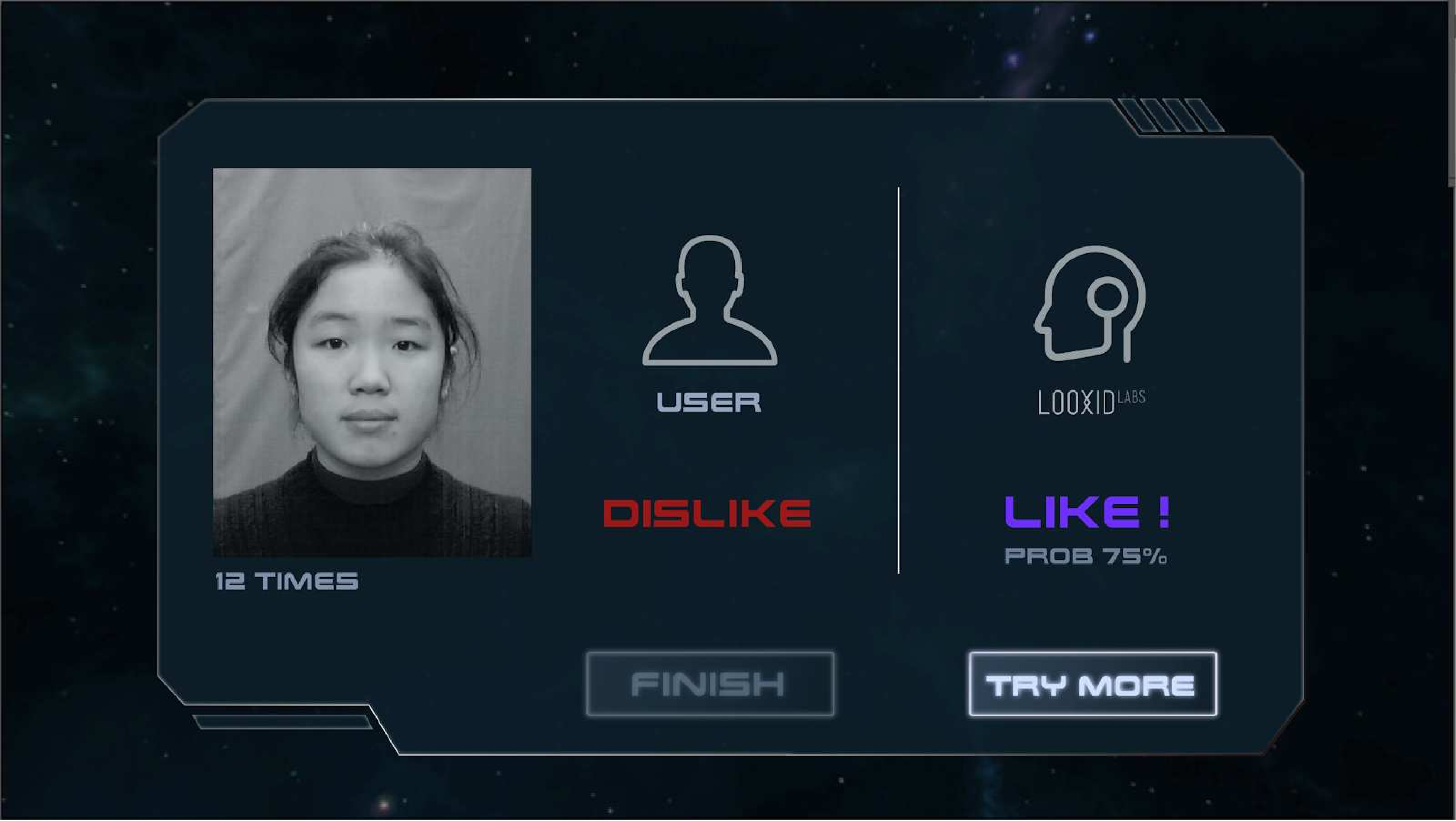}
\centering
\caption{Screenshot of a comparison between a user selection and the result of the attractiveness recognition algorithm.
}
\label{fig_vive_result}
\end{figure}
The second session was conducted to assess the rating of the image, and then to immediately compare the system's prediction with respect to the facial attractiveness. With respect to the real-time attractiveness recognition system, the application consists of two phases: (1) training and (2) prediction. In the training phase, participants looked at 30 different human face images randomly selected from the image set of a total of 200 pictures that are publicly available on the web, and they evaluated those images individually by simply answering whether he or she likes or dislikes each image. For the prediction phase, we chose the pre-defined set of features for which the performance was proven by our previous experiments to present the result of our attractiveness recognition algorithm in real-time. Specifically, HOG (EEG feature) and fixation frequency (eye feature) were selected, and the polynomial kernel order of the SVM was set to 4. During a total of 10 trials of the prediction comparison result phase, we adopted the incremental SVM classifier so that at every third, sixth, and ninth trial, the SVM classifier is updated to include all of the newly acquired data. The system provided a result with the posterior probability value as well as the participant's selection, as shown in Figure~\ref{fig_vive_result}.

The survey, which was conducted after the VR session, was designed to later  acquire feedback about the system. The questions mainly aimed to determine their opinions and concerns about a system that could interact with users by understanding their emotions.

\section{RESULTS}
\subsection{System accuracy}
We firstly verified the accuracy of the attractiveness recognition algorithm. A 10-fold cross-validation method was applied to the data acquired from the first session of the experiment. For example, 27 images were randomly selected and used as a training set, and the remaining three images were used as a test set for face, clothes, and color images. We used 49 randomly selected images as a training set, and the remaining five images were used as a test set for composite images. A model created using the training set was applied to the test set to calculate the posterior probability of the degree to which the participant is attracted to an image in the test set. If the probability exceeds 0.5, we determine that the participant is attracted to the image, and if it is lower than 0.5, the image is not attractive. The prediction results were compared with the user's assessments of the image to analyze the accuracy of our proposed algorithm. The mean and standard deviation of the accuracy obtained by repeating the above process 10 times are shown in Table~\ref{accuracy}. 

In general, the results were statistically significant in that the system recognized the participants' attraction. The accuracy of the composite images was 0.744, which was the highest, and that of the color images was 0.605, which was the lowest. While we tried to organize data into which people could clearly feel an attraction for face, color, and composite images, it was difficult to make clear to feel attracted in the case of color images. We found higher system accuracies with datasets that had more robust attraction. We present a more in-depth analysis of this topic section 7.

We checked one more result during the process of determining the system accuracy. It is known that the uniformity and robustness of the data to be learned are important for the machine-learning model. Thus, there were inferences that the same type of visual stimuli should be used. To consider the hypothesis, we operated the system utilizing all 144 images, including face, clothes, color, and composite images. As with the above analysis process, we used a randomly selected proportion of 90\%, i.e., 130 images, as a training set, and the remaining 14 images as a test set. The accuracy is lower than the result obtained when using only the composite images, but it is higher than the result obtained when using only face, clothes, and color images, as shown in the last graph in Table 2. We hypothesize that this is because the system recognizes attractiveness using biometric data when attraction occurs regardless of the type of visual stimuli.

\begin{figure}[t]
   \begin{subfigure}[b]{0.45\linewidth}
   	\centering 
   	\includegraphics[width=0.95\linewidth]{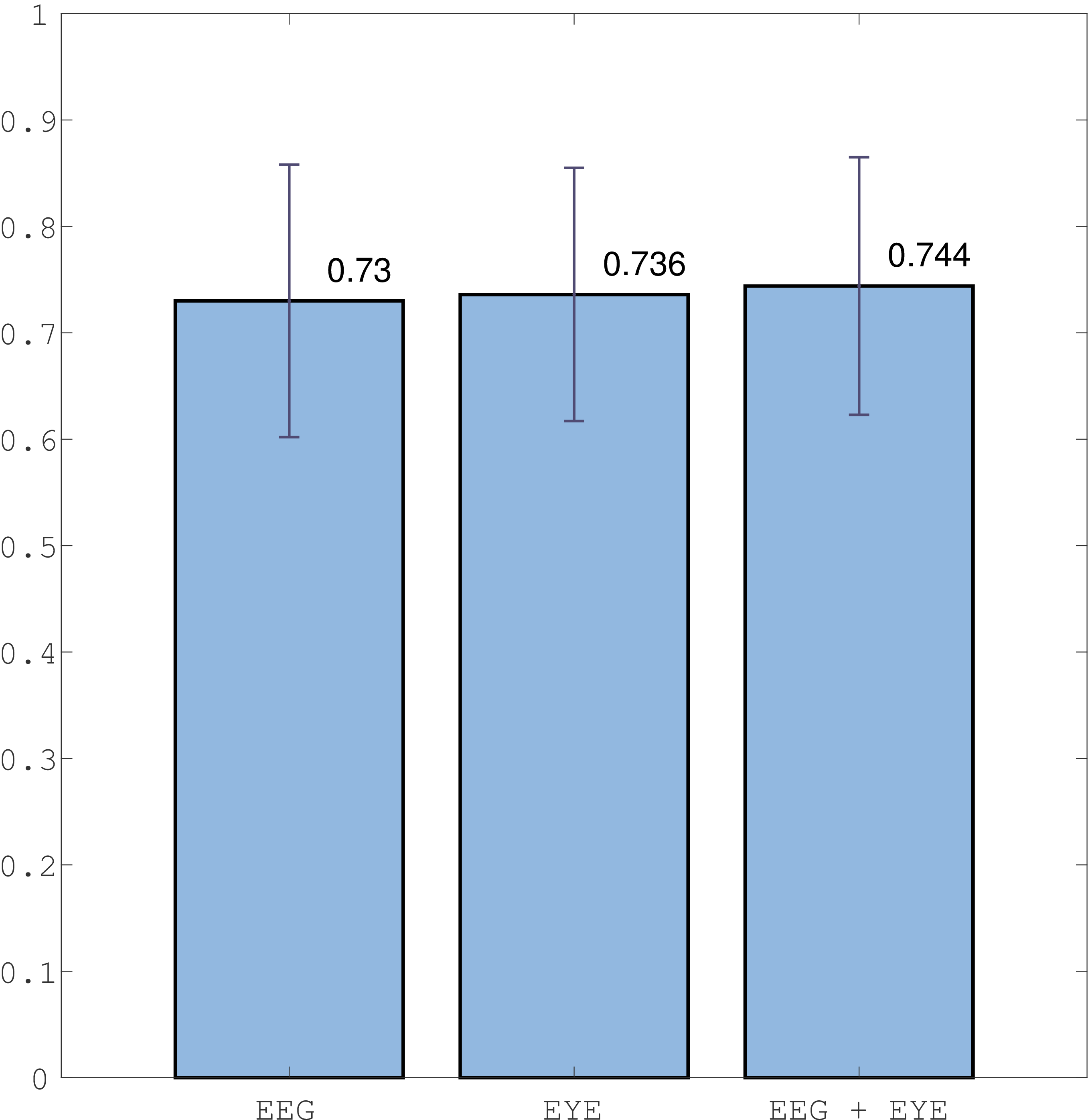}
   	\caption{} 
\end{subfigure}%
\begin{subfigure}[b]{0.45\linewidth}
   \centering 
   \includegraphics[width=0.95\linewidth]{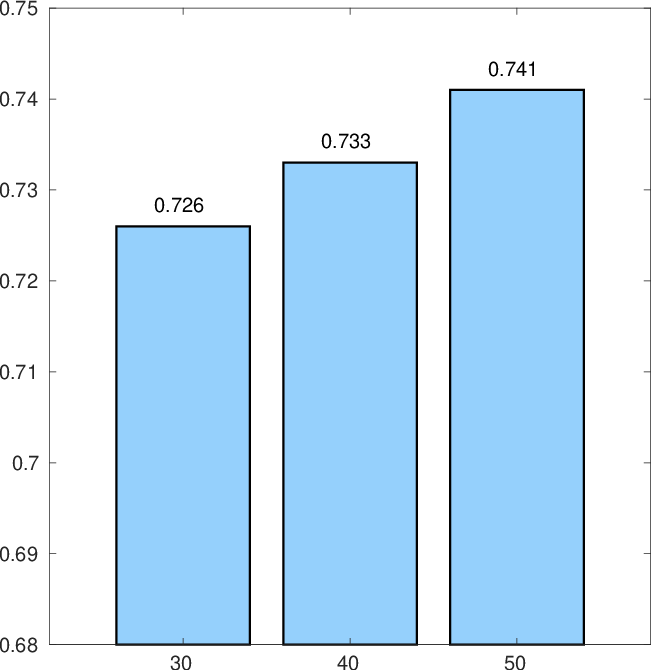}
   \caption{} 
\end{subfigure}%
   \caption{Comparison of accuracy using (a) different sensor modality and (b) different sized training set (right).} 
   \label{fig_acc}
\end{figure}

We performed additional analysis on factors that could affect the system accuracy using data on the composite images that had the highest accuracy. Figure~\ref{fig_acc} shows a comparison of the accuracy for the combination photos between different sensor modalities. The accuracy when both the EEG and eye data are processed is higher than when only one of the two is used. In other words, in terms of prediction accuracy, it is better to design the attractiveness recognition model to be multi-modal. 

\begin{table}
  \caption{System accuracy for each dataset.}
  \label{accuracy}
  \begin{tabular}{ccccc}
    \toprule
    Face & Cloth & Color & Composite & All \\
    \midrule    
	0.644 & 0.645 & 0.605 & 0.744 & 0.692 \\
  \bottomrule
\end{tabular}
\end{table}

From the experiment, we further investigated a way of improving the accuracy for our future work. Within our acquired dataset, we tested whether the accuracy changes as the size of the training data increases. As the characteristics of the stimuli should be constant to make a justifiable comparison, the combination image set was chosen for the additional analysis. As illustrated in Figure~\ref{fig_acc}, the difference is not sufficiently significant to explicitly state that the greater the accuracy, the larger the size of the training set. Nevertheless, we would continue a similar investigation to improve the system performance.

\subsection{Example of practical use: Factor analysis}
As one example of how the system could be put into practice, we performed a factor analysis on composite images to demonstrate the feasibility of the proposed system. Several studies have been conducted to analyze the factors that affect users when rating images or videos. For example, if someone gave 4-stars on a YouTube video after watching, marketers would seek to obtain information about which scene or which item had the greatest impact on this judgment. The most representative method is the gaze-point analysis to assess how long factors have been watched using eye-tracking technology. In this study, we compared the gaze-point analysis and the correlation analysis using the results of our attractiveness recognition algorithm for the three components of the composite image, namely the face, clothes, and background color.

\begin{figure}[!t]
\includegraphics[width=\linewidth]{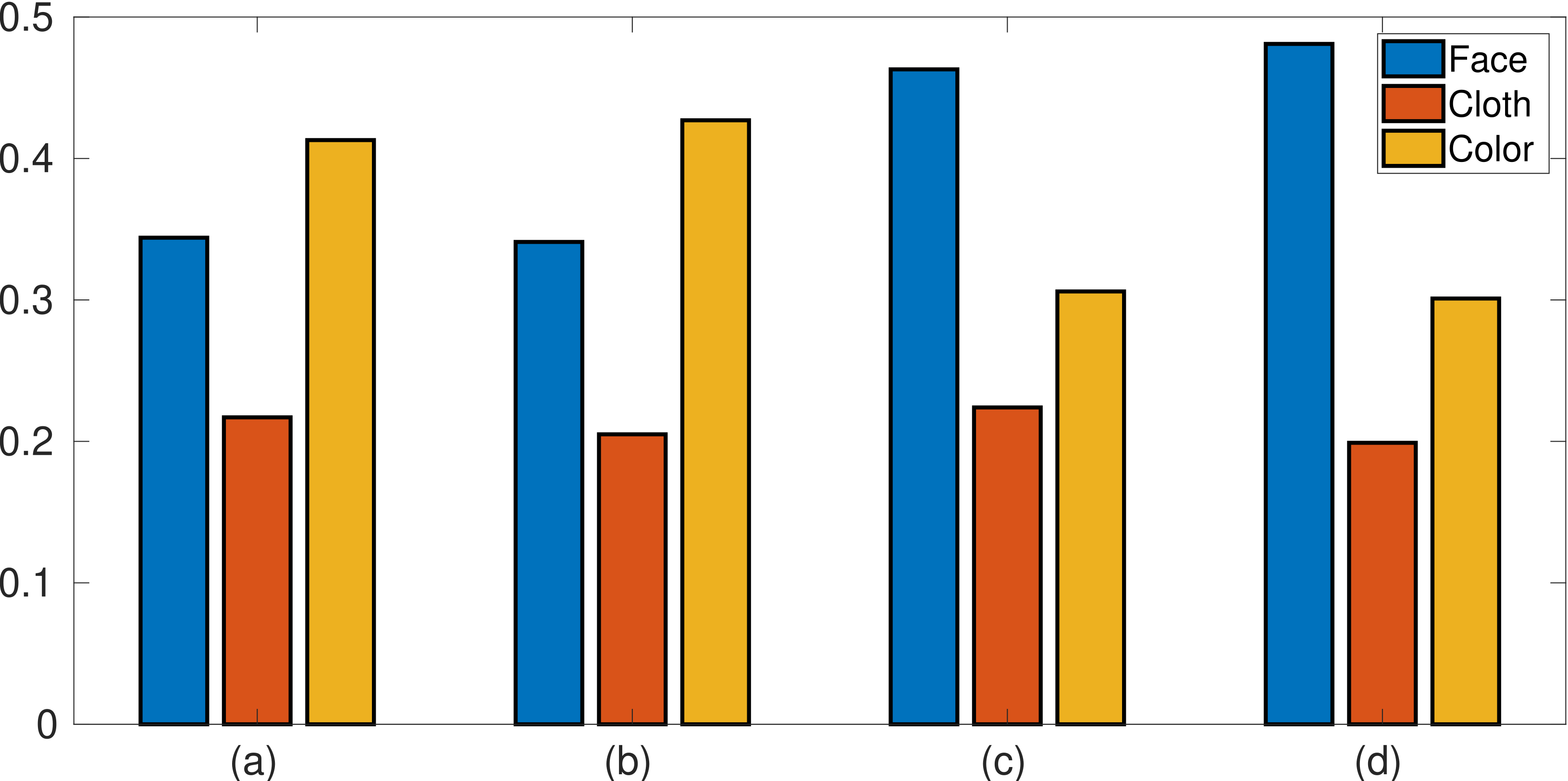}
\caption{Gaze ratio of face, clothes, and color areas in the composite image (a) Male watching male images (b) Male watching female images (c) Female watching male images (d) Female watching female images.
}
\label{fig_gaze_ratio}
\end{figure}

\begin{figure}[!t]
\includegraphics[width=\linewidth]{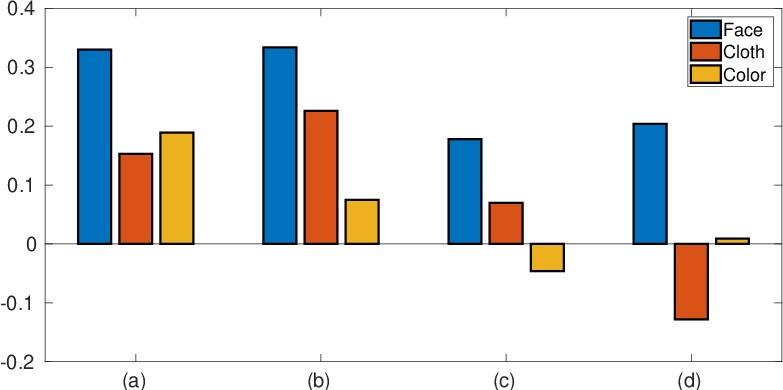}
\caption{Pearson correlation between predicted attraction on the composite image and on face, clothes, and color images comprising the image (a) Male watching male images (b) Male watching female images (c) Female watching male images (d) Female watching female images.}
\label{fig_corr}
\end{figure}

We set the face, clothes, and background color areas in the composite image to different regions of interests (ROIs), and we obtained the amount of gaze data with a sampling rate of 60 Hz, which means that 60 gaze points were collected per second in each ROI. Then, in Figure~\ref{fig_gaze_ratio}, we compared the results for male and female participants when watching images of men and women, respectively. The gaze ratio of the background color was the highest for male participants, while the female participants watched the face area the longest. With these results, it remains unknown whether we can assume that the background color had the greatest influence on men, while the face had the greatest influence on women.

Next, we analyzed the correlation between the attraction on the composite image and that on each factor of the image using our proposed algorithm. We calculated the posterior probability of attractiveness for every face, clothes, color, and composite image through a model built using all 144 images. We obtained the Pearson correlation of the probability results for all images, including between the composite images and the respective face images, clothes images, and background color images that made up the composites. As with gaze-point analysis, we compared the results of male and female participants when they viewed male and female images, respectively, as shown in Figure~\ref{fig_corr}. In all cases, the image of the faces had the highest Pearson correlation, and this was shown to be higher overall for males than for female participants. Only the facial factor showed a moderate positive relationship in the results for male participants, and all other factors showed a weak relationship.

The results of the correlation analysis using our proposed system are more reasonable than the results of the gaze-point analysis compared with the previous studies on factors that men and women consider important when evaluating the attraction of others. It shows the limitation of being difficult to understand the intent of why participants look at the point using only eye-tracking technology. However, the attractiveness recognition model using biometric data can be expected to be more useful in finding the most important factors in an image or video.

\subsection{Participant interview}
Participants gave positive feedback on the system after experiencing the interaction that shows the results of the system in real-time in the second session. They believed that the system was more accurate and efficient than other emotion recognition technologies because they recognized our system utilizes directly brain activity even though they didn't understand how the algorithm works.
\begin{quote}
\textit{"I know that emotion is a result of diverse activities. Therefore, it makes sense to classify emotion by different brain wave patterns."}
\end{quote}

We engaged in brainstorming to hear various opinions from participants regarding how they wish for our system to be utilized. Many thoughts were shared, including a content recommendation system, dating application, entertainment utilities, education assistive technology system, mental-care solution, crime analysis, virtual counsellor or friend, among others.
\begin{quote}
\textit{"I can enjoy my mealtime if I can talk to AI who definitely knows a lot about me. I would feel less lonely being together with it. Like movie Her"}
\end{quote}
\begin{quote}
\textit{"The system will be able to recommend my ideal type in the dating application."}
\end{quote}

However, some persons were doubtful about the possibility and also expressed some concerns. Therefore, the system should be developed to not only meet persons' expectations, but also to prevent any possible side effects such as invasion of privacy or deprivation of human roles.
\begin{quote}
\textit{"If individual attraction becomes one of personal information, some people will feel uncomfortable."}
\end{quote}

\section{IN-DEPTH DISCUSSION WITH EXPERTS}
In our approach, we combined multidisciplinary technologies such as neuroscience, biology, data science, machine learning and human interaction. Therefore, the professional knowledge from diverse fields are indispensable to evaluate the emotion recognition system and discuss its applicability and capabilities. Therefore, we had in-depth talks with four experts from different domains.

During an individual interview lasted for about an hour or up to four hours, we first talked about their expertise and the significance of the emotion recognition interface in their field. Second, we briefly introduced our system framework and results of both the experiment and the preliminary user interaction study, and they provided feedback about those. Finally, we led an intensive discussion about the future work of the system. Throughout the interview, very much meaningful feedback was exchanged, and interesting as well as practical directions in which the system should be developed and applied were discussed.

Four experts included a neuroscientist (N), an AI research scientist (A), a VR movie director (D) and a CEO of a startup developing an emotional robot (E). All of them are disinterested parties of our work, so we had them indirectly experience the system. All of them have shown their proficiency on their job.

\subsection{Evaluation of the system and experiment}
The experts evaluated the system and the experiment, and proposed how they could be improved.
\subsubsection{System framework}
In recent years, EEG-based interfaces have had different shapes and incorporated a number of sensors. N mentioned that the emotion recognition depends on the interface specifications. With respect to our system, it is highly promising that both EEG and eye tracking are used.
\begin{quote}
\textit{"The eyes, in particular, contains much significant information such as the user's interest and concentration. Thus it is one of the most actively studied subjects in the field of neuro-biology." (N)}
\end{quote}

Owing mainly to hardware issues, the measurement accuracy of the eye gaze remains insufficient. In this system, better post-processing would be possible when eye movement information is analyzed together with EEG. In the case of the EEG, the sensor position and quantity are critical to ensure accurate data capture and analysis. However, bulky EEG equipment has less usability. Therefore, he appreciated the placement of sensors in our system from the perspective of the efficiency of the position relative to the prefrontal region.
\begin{quote}
\textit{"As each brain region is not clearly divided according to different functions, in the case of EEG, it would be best to capture brainwave data from the whole hemisphere. In this system, focusing on prefrontal region was a good decision as every information in our brain is collected and processed in the frontal lobe at the end." (N)}
\end{quote}
 
\subsubsection{Prediction model and experimental design}
It is recognized that data selection is very important when developing a prediction model. Strictly classified images could improve the system validity. Therefore, stimuli should be chosen very carefully. However, in our experimental design, for more general use outside the laboratory, we used three kinds of stimuli, namely face, cloth, color, and their combination. These stimuli are related to both the biological attraction and desire of purchase. More complicated biological reactions can be included compared with signals with simple stimuli, and effective learning may be more difficult. However, as mentioned by N with respect to scene processing, if our system is sufficiently accurate, the method could be efficiently adaptive in many research fields.
\begin{quote}
\textit{"About the experiment design itself, it seems like the images contained a bit too complex information. However, what is most interesting in this experiment is that it tracked individual's scene processing along with corresponding physiological signals. The observation of reciprocal interactions between contents, attraction level, eyes and brain activities are itself extremely meaningful. Therefore, if it is revealed that an accurate measurement of human's attraction on complex objects through machine learning is possible, the study will have a huge significance in the field of neuroscience." (N)}
\end{quote}

Our next question pertains to the standard having sufficient accuracy. The AI research scientist referred to the amount of data. A lot more data should be acquired to obtain more meaningful information.
\begin{quote}
\textit{"The number of accumulated data has to be big enough to mathematically explain how the system has improved through a continuous learning." (A)}
\end{quote}
This means that it is necessary to prove what is defined as an efficient accuracy with more data by performing experiments for the next step. In addition, if our proposed system is designed to offer individualized service, it is important to encourage users to keep their answers until the system reaches an optimal training session point.

\subsection{Need for continuous and seamless sensing}
The experts reported that our proposed system is well suited to overcome limitations of conventional attractiveness evaluation methodologies from the perspectives of continuous and seamless measurement. In a recommendation system where persons' attraction information is very important, the users' action log and self-assessment result work as the main guideline in building up a prediction model. However, the conventional method fails to obtain detailed information about the various elements that are included in the scene. It is also necessary to force the users to evaluate many things. Recently, there have been attempts to read users' emotions by investigating their facial expressions or voice tones. However, it also has limitations in terms of detecting more complex emotions. Experts confirmed the significance and applicability of our proposed system in that it allows an unobtrusive and detailed measurement of users' reactions and attraction to various content.

\begin{quote}
\textit{"Users generally skip answering thus it is hard to collect enough data. I think the best thing about this system is that it allows a seamless measurement of users' attraction on even a very specific part of the contents without bothering them." (A)}
\end{quote}
\begin{quote}
\textit{"Facial expression, in particular, is a clear, universal and consistent method which can easily be sensed with a camera. However, more complex and subtle emotions are not caught in facial expressions, and it is hard to acquire enough data to classify those according to attractiveness of varying objects. Therefore, collecting physiological data using a VR platform would be a better in conducting lots of different experiments." (D)}
\end{quote}

\subsection{Expectation of emotion recognition interface}
In the preliminary user interaction study, general users' various ideas about how they wish our system to be used were shared. Each of the professionals elaborated on those ideas based on their expertise, suggesting many more concrete application scenarios. Even though they are from varying fields, their stories were closely related based on the goal of the system: to realize an interactive design that combines attractiveness evaluation with corresponding cognitive changes.

\subsubsection{Interactive design}
The experts all agreed that the application of an emotion recognition interface should be useful for designing a strong emotional connection between humans and technology. Emotional connection with characters is very important in VR utilities, scenario design, game design, and a virtual assistant because it will help the users to become more immersed. In a VR-based movie, for the main characters to sympathize with the user and react properly, an attractive recognition model is required.
\begin{quote}
\textit{"Recently, interactive storytelling techniques based on the dynamic object structure, where the scenario goes differently according to user's certain response, are receiving much attention. If this system is adapted to the interactive storytelling and controls the dynamic path in the multi scenario structure, we can provide a more personalized and immersive experience." (D)}
\end{quote}
In the case where a voice agent is adapted in VR, if it is to be designed to take good care of the users, it is very important to develop a system that correctly recognizes their moods and reacts to jokes or surroundings. Furthermore, in the study of robot interaction, tests involving a prototype of the VR robot are useful for researching human reaction.
\begin{quote}
\textit{"I think the system can do much to achieve interaction design. Even if it is not about VR, any kinds of robot-human interactions will be much easily tested with your system. The analysis of the users' attraction then can be applied when making a real robot." (E)}
\end{quote}

\section{DISCUSSION}
In the first session of the experiment, we measured the accuracy of our attractiveness recognition model. The experts evaluated the system as being sufficiently accurate. In general, the loss of data should be considered negligible in order to improve wearability of the headset. Major factors that led to favorable results include a sensing system that can reliably measure bio-signals, a time-synchronization algorithm and an attractiveness recognition algorithm that utilizes both brain engineering and machine learning technologies. 

One of the greatest achievements found in the accuracy analysis process was that the results were shown without combining the different types of visual stimuli. Many neuroscience studies have combined the dataset into the same type when designing experiments, fearing that using different types of datasets will activate different areas of the brain~\cite{ochsner2008cognitive}. However, the results show that the biometric pattern does not vary much for different types of datasets when the user experiences attraction. If further research proves this hypothesis, our interface can recognize the user's attraction with only the bio-signals that we measured, regardless of the type of stimuli in any environment.

The experiment results provided a number of ways we can improve the accuracy of an algorithm. First, more accurate results can be obtained when using a data set that has a clear attraction to the user. When we look at the accuracy according to the type of dataset, the accuracy is the lowest for the types of data that people typically don't clearly distinguish their likes and dislikes (e.g. color) . With data types with higher accuracy (e.g. face, clothes) using a composite image, where all elements were added and people's tastes could be reliably reflected, led to the highest accuracy. It is expected that when a person likes an object,  there are clear characteristics of the bio-signal that indicate attraction. Therefore, well-designed data collection will help improve accuracy. In addition, if we add a category such as "neutral" instead of simply discriminating whether a person likes or dislikes stimuli, a more reasonable system will be designed. The second factor is the usefulness of the multi-modal system. The accuracy associated with simultaneously using both EEG and eye-tracking data was higher than the accuracy of using data acquired from a single modality, but the difference was not significant. There is a lack of proof of our hypothesis that the more abundant the data, the better the results. Further studies are needed to develop a learning model that can generate synergies when combining various sensor modalities. Third, as with the modality comparison, the accuracy did not increase significantly as the sample size of the training set increased. When implementing a general machine learning model, the the larger number of data samples had better performance~\cite{friedman2001elements}. However, increasing the number of data samples by a factor of 10 did not result in significantly improved results in each case. In future work, we need to see how much improvement in model accuracy can be achieved by using more data.


The results of the factor analysis showed that this system has a variety of possibilities for practical use. When a user observes a car design in a VR environment, the factor analysis results can be used to determine which elements of the car have had a significant impact on the car's attractiveness or the user's purchase decision. We can investigate what factors affect the user's attraction while they experience the space such as a retail store or an airport built in a VR environment. However, in this study, factor analysis was not performed in one image owing to limitations of the model currently being developed. In a strict sense, factor analysis should compare factors in one image as in the gaze-point analysis. However, it is not currently possible to perform strict factor analysis using our algorithm because it recognizes the attractiveness of the entire visual stimuli being watched by the user. Therefore, in this study, we obtained the correlation using the obtained results after showing the factors that make up a single image. We plan to develop a model that can recognize attractiveness for each factor within a single scene. We have the capabilities to develop an advanced model that allows us to analyze how much attraction was felt when observing specific factors because eye-tracking technology is employed in addition to the attractiveness recognition algorithm.

In interviews with participants and experts, there was some concern about privacy, but overall, there are many expectations about the use of this system. There are many interesting ideas about potential uses, such as psychological analysis, criminal analysis, and utilization in dating services. Among them, examples of applying the emotion recognition interface to AI have led to the expectation that innovative HCI can be developed. If AI that is capable of speaking in a manner similar to that of Apple's Siri can recognize the users' emotions, the world of the film "Her" could become a reality. Because the VR environment has characteristics that are different from existing digital environments, AI technology and interactive storytelling can be actively utilized. In this environment, if the AI reacts or the story of the contents changes based on the attractiveness, a more immersive VR world can be built. Our proposed system will contribute to the HCI field as a new interface that emotionally connects a user and computer in these diverse applications.

\section{CONCLUSION}
Our proposed emotion recognition interface showed high accuracy with seamless sensors in the experiment and interviews. We built a continuously sensing system using the characteristics of HMDs, and developed a machine-learning model that can recognize the user's attraction using the data obtained from the sensors. The system has two advantages because it utilizes biometric data. First, the genuine emotion can be recognized by analyzing the bio-response that occurs instantly as the user's emotion changes. Second, the system continues to measure bio-signals and monitor the attractiveness information in a time-series while the user utilizes the content.

This interface has good potential for use various applications, such as in the various cases presented by experts in our interviews. Using the interface, a computer understands which object, character, or item a user is interested in, and recommends personalized content, provides interactive storytelling, or responds through AI. The proposed system is expected to be used as a new tool for HCI by improving the accuracy and advancing algorithms in further studies.

\balance{}

\bibliographystyle{SIGCHI-Reference-Format}
\bibliography{ref}

\end{document}